\documentclass[english,aps,prl,twocolumn]{revtex4-1}
\usepackage[T1]{fontenc}
\usepackage[latin9]{inputenc}
\usepackage{graphicx}
\usepackage{babel}
\usepackage{amssymb}
\begin{document}

\title{Classical simulability of noisy boson sampling}

\author{J.J. Renema\textsuperscript{1)}{*}, V. Shchesnovich\textsuperscript{2)},
R. Garcia-Patron\textsuperscript{3)}}

\affiliation{1) Complex Photonic Systems (COPS), Mesa+ Institute for Nanotechnology,
University of Twente, PO Box 217, 7500 AE Enschede, The Netherlands,}
\email{j.j.renema@utwente.nl}
\affiliation{2) Centro de Ci\^{e}ncias Naturais e Humanas, Universidade Federal do
ABC, Santo Andr\'{e}, SP, 09210-170 Brazil.}
\affiliation{3) Centre for Quantum Information and Communication, Ecole Polytechnique
de Bruxelles, CP 165, Universit\'{e} Libre de Bruxelles, 1050 Brussels, Belgium.}

\begin{abstract}
Quantum mechanics promises computational powers beyond the reach of classical computers. Current technology is on the brink of an experimental demonstration of the superior power of quantum computation compared to classical devices. For such a demonstration to be meaningful, experimental noise must not affect the computational power of the device; this occurs when a classical algorithm can use the noise to simulate the quantum system. 

In this work, we demonstrate an algorithm which simulates boson sampling, a quantum advantage demonstration based on many-body quantum interference of indistinguishable bosons, in the presence of optical loss. Finding the level of noise where this approximation becomes efficient lets us map out the maximum level of imperfections at which it is still possible to demonstrate a quantum advantage. We show that current photonic technology falls short of this benchmark. These results call into question the suitability of boson sampling as a quantum advantage demonstration.
\end{abstract}

\maketitle


In recent years, interest in quantum information processing has focused
onto small-scale quantum computing machines, which could perform specific
tasks of scientific or technological interest faster than classical
computers, and which can be constructed with current or near-future
technology \cite{Wolf2018,Preskill2018}. An important milestone would be demonstrating a quantum device which can convincingly outperform a classical computer at some particular problem, i.e., a demonstration of a quantum advantage \cite{Montanaro2017, Aaronson2011}. For most quantum advantage proposals, the required system size is predicted to lie around 50 particles/qubits \cite{Boixo2018b, Wusupercomputer, Pednault2017}, and constructing a quantum system of that size is the current focus of experimental quantum information efforts worldwide \cite{Kellly2019, Gambetta2019, Zhong2018, Gong2018, Bernien2017, Friis2017}.

A major hurdle to experimentally achieve a quantum advantage is our poor understanding of how noise affects the viability of quantum advantage demonstrations \cite{Chen2018a, Boixo2018a, Boixo2018b, Bremner2017, Markov2018, Villalonga2018, Gao2018, Yung2017, Patron2017, Oszmaniec2018, RahimiKeshari2016, Renema2018,Shchesnovich2014,Arkhipov15,Aaronsonweb,Lund2014,Aaronsonlostphotons}. Since near-term quantum devices are not error corrected, it is generally believed that they are not scalable to arbitrary size, succumbing to noise for a sufficiently large system size or a sufficiently high level of noise. The challenge lies in finding the point where this occurs for any given system.

The transition from quantum to classical is demarcated by the point at which a classical algorithm can use the inevitably present experimental noise to simulate the task being performed by the quantum system. For such an algorithm to meaningfully restrict the demonstration of a quantum advantage in a given system, it must meet three criteria. First, it must use the types of noise naturally present in the system, at the levels at which they are present in experiments. Second, there must be a well-defined point in terms of system size and noise level where the algorithm becomes efficient; this point then demarcates the level of imperfections admissible in that experiment. Finally, in order for the algorithm to be applicable to near-term devices, the point where the algorithm becomes efficient must lie below the system size being targeted by quantum advantage demonstrations.

In this work, we show an algorithm to classically simulate one quantum advantage platform, namely boson sampling (interference of noninteracting, indistinguishable bosons) \cite{Aaronson2011}, in the presence of the two major sources of noise for that experiment, namely particle loss and distinguishability. Our algorithm meets the three criteria outlined above. It efficiently simulates interference of $n$ noisy bosons as polynomially many  interference processes of size $k \leq n$, supplemented with $n-k$ classical bosons. Remarkably, $k$ only depends on parameters which quantify the level of noise, and not on the number of particles undergoing interference. This shows that boson sampling at a given level of imperfections only carries the hardness of $k$ interfering bosons, where $k$ acts as an upper bound on the size of a boson sampler that can be constructed in the presence of a given level of noise. Such a $k$ exists for any level of noise, which shows that noisy boson sampling is asymptotically non-scalable. Our results imply that experimental noise defines the maximum size to which a noisy boson sampler can be scaled. Our algorithm is efficient for systems constructed with the best photonic devices currently available.



In boson sampling, the task is to provide samples of the output probability distribution resulting from many-boson interference. Aaronson and Arkhipov \cite{Aaronson2011}
provided strong complexity-theoretic evidence that this task cannot be simulated efficiently
(i.e. in polynomial time) on a classical computer. Boson sampling was initially proposed in the framework of linear quantum optics, but alternative implementations for ion traps \cite{IonBS}, one-dimensional optical lattices \cite{OLBS} and superconducting qubits \cite{SCBS} already exist. Moreover, applications for boson sampling devices have been proposed, particularly in quantum chemistry \cite{Huh} and machine learning \cite{Xanadu}. The point where a boson sampler is expected to outperform a supercomputer is 
around 50 bosons \cite{Wusupercomputer}, an observation that has spurred
a range of experimental efforts \cite{Spring2012,Broome2012,Tillmann2013,Crespi2013,Bentivegna2015,Carolan2015,Wang2017, Wang2018, Zhong2018}.

Progress has also been made on understanding the effect of imperfections on boson sampling. The two main imperfections in boson sampling are photon loss, where some bosons are lost in the course of the experiment, and distinguishability, i.e. bosons having different internal quantum states. Boson sampling with partially distinguishable photons was shown to be classically simulable in \cite{Renema2018}. For loss, two sets of results can be singled out. First, boson sampling was shown to retain its hardness \cite{Shchesnovich2014, Arkhipov15, Aaronsonweb,Lund2014,Aaronsonlostphotons} for small imperfections, including loss of a constant number of photons (i.e. where the survival probability of each photon goes to one asymptotically). Secondly, boson sampling was shown to be classically simulable when the loss probability increases with the number of photons (i.e. where the per-photon survival probability goes to zero asymptotically) \cite{Patron2017,Oszmaniec2018}. The experimentally crucial case of a constant surival probability per photon (also known as {\em linear loss}) remained unsolved, and was identified as one of the major open problems in boson sampling \cite{Aaronsonweb2,Aaronsonlostphotons,Wang2018}.

We demonstrate a classical simulation algorithm for many-body bosonic interference where the per-boson transmission probability $\eta$ is fixed and where the bosons may also have some degree of distinguishability $x$, where $x$ is the internal wavefunction overlap between two arbitrary bosons. We show that for every $x < 1$ and $\eta < 1$ (i.e. for any level of noise), boson sampling can be approximated as quantum interference of clusters of only $k$ particles. Since $k$ does not depend on the number of bosons $n$, it acts as an upper bound on the size of boson sampler which can be constructed with components of given transmission and bosons with given distinguishability. Since the hardness of a boson sampler is fixed by the number of interfering bosons, and since such an upper bound $k$ exists for any level of noise, boson sampling is asymptotically non-scalable. Our work answers the question of whether boson sampling with linear losses is classically simulable in the affirmative. 

We can use our results to estimate the quality of experimental components required to demonstrate a quantum advantage. We find that a transmission of $\eta>0.88$ is necessary to simulate boson sampling with 50 bosons at an accuracy level of 10\%, and that the current best boson sampling platforms  are restricted to interference of 21 bosons under the same criterion. This shows that achieving a demonstration of a quantum advantage with boson sampling requires more than the construction of high-rate, large-scale photonic systems, as was believed previously \cite{Neville2017}: it also requires a qualitative improvement in the equipment used.

\begin{figure}
\includegraphics[width=5cm]{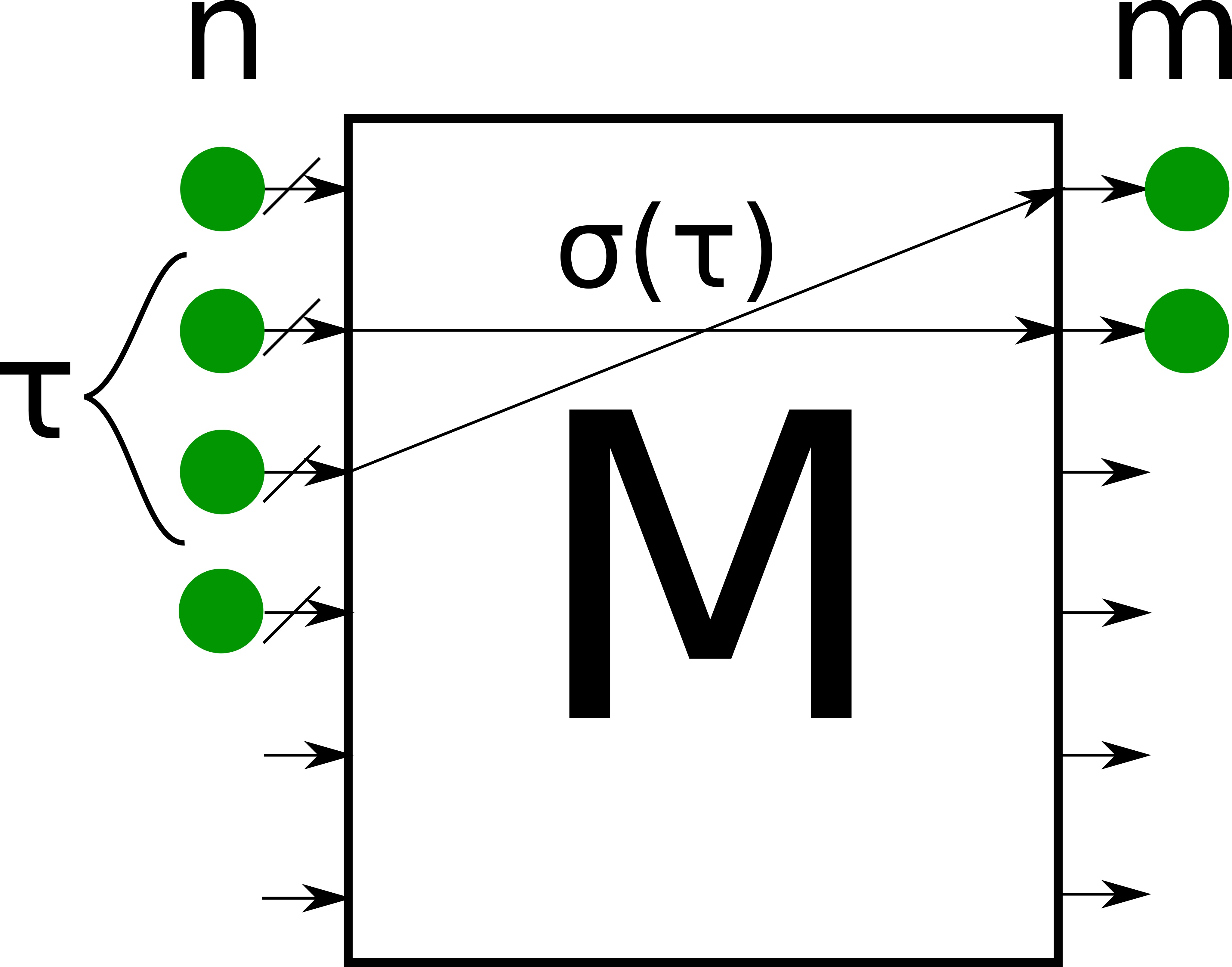}\caption{A pictorial representation of a boson sampler with losses. We consider a system with $n$ input particles and $m$ particles at the output, where the particles are lost before they enter the interferometer. The symbols illustrate the notation used in this article: $\tau$ are all ways of selecting $m$ particles from $n$, and $\sigma(\tau)$ are all the ways of permuting those particles. In the figure, we give an example of both.}
\end{figure}

We begin by setting up the problem, see Figure 1. The initial boson sampling proposal concerns the interference of  $n$ single-boson input state
over an $N$-mode coupling interferometer modeled by a unitary transformation $U$.
At the output we measure the particle number in each mode.
A condition for the hardness proof to hold is that
the number of modes $N$ obeys $N=O\left(n^{2}\right)$,
which guarantees that the probability of two bosons emerging from the same mode (known as a collision)
can be neglected.
In that case, the probability
of the photons exiting the system in a particular set of modes is given by $P=|\mathrm{Perm}(M)|^{2}$ \cite{Scheel2008},
where $M$ is a submatrix of the overall unitary transformation $U$
that is chosen by selecting the rows and columns corresponding to
the input and output modes respectively, and $\mathrm{Perm}(M)=\sum_{\sigma}\prod_{i}M_{\sigma_{i},i}$
is the permanent function \cite{Minc1978}, where $\sigma$ is a permutation
and the sum runs over all permutations.

The hardness of boson sampling ultimately stems from the fact that
for an arbitrary matrix, the permanent cannot be approximated efficiently
by a classical computer \cite{Valiant1979}. The best known algorithms
for computing an arbitrary permanent are due to Ryser and Glynn \cite{Ryser1963,Glinn2010}.
These algorithms scale as $p2^{p},$ where $p$ is the size of the
matrix. It was shown recently \cite{Neville2017} that by using a
Metropolis algorithm, one can generate samples from a
probability distribution where each entry is given by a
permanent, at the cost of evaluating a constant number
of permanents. The hardness of boson sampling therefore
rests on the hardness of computing the individual
output probabilities of a boson sampler.

In this paper, we solve how this hardness is compromised by loss and
distinguishability. If the losses in each
path of the boson sampler are equal, which is a reasonable approximation
in experiments, the action of the interferometer is equivalent to 
a circuit where all losses act at the front of the experiment
followed by an ideal interferometer $M$
\cite{Neville2017,Oszmaniec2018,Aaronsonlostphotons}. 
The stochastic nature of the losses will make the number of transmitted photons $m$ fluctuate
according to a binomial distribution, but
for the purpose of keeping the presentation simple we first present 
an algorithm for fixed pair $m$ and $n$ and return later to the 
analysis of the most general case. This scenario fits
a recently suggested proposal to circumvent losses in boson sampling 
by enforcing specific combinations of $n$ and $m$ by post-selecting, 
as was done recently for $m=5,$ $n=7$ \cite{Wang2018}. Our result 
strongly constrains the viability of this post-selection approach.

To derive our results, we will consider the probability $P$ of an arbitrary collisionless output configuration, without loss of generality. When only $m$ bosons are detected out of the initial $n$,
the detection probability at the output results from the incoherent sum 
over the $\left(\begin{array}{c}
n\\
m
\end{array}\right)$ different ideal boson sampling terms, \begin{equation}
P={n\choose m}^{-1}\sum_{\tau}|\mathrm{Perm}(M_{\tau})|^{2}={n\choose m}^{-1}\sum_{\tau} P_{\tau},
\end{equation}
where $\tau$ is an $m$-combination of $n$ and the sum runs over
all such combinations. Each $P_{\tau}$ is the probability corresponding
to lossless boson sampling if $m$ bosons were injected in the modes $\tau$. The prefactor ${n\choose m}^{-1}$ arises because of normalization.

Our strategy is to break
up equation (1) into terms which correspond to classical transmission,
two-boson interference, three-boson interference, and so on. We
will then show that boson losses reduce the weight of the higher
interference terms. The consequence of this is that beyond some number $k,$
which is only a function of $\eta,$ these terms can be neglected. We can then use our approximation as the input for a
Metropolis sampler, which can sample efficiently from
our approximate probability distribution.

Our starting point is the following expression for the detection probability
in the lossless, fully indistinguishable case \cite{Tichypartial}:
\begin{equation}
P_{\tau}=|\mathrm{Perm}(M_{\tau})|^{2}=\sum_{\sigma}\mathrm{Perm}(M_{\tau,1} \circ M_{\sigma(\tau),1}^{*}),
\end{equation}
where $\sigma$ is a permutation of the elements of $\tau$, the sum
runs over all permutations, $\circ$ denotes the elementwise product,
$^{*}$ denotes complex conjugation, and $M_{\sigma,1}$ denotes permuting
the rows of matrix $M$ according to $\sigma$ and the columns according
to the identity. We will use this notation throughout. For the purpose
of keeping the mathematics simple, we shall derive our results for
perfect indistinguishability,
reintroducing distinguishability at the very end.

Equation 2 can be rewritten by grouping terms according to the number
of fixed points (unpermuted elements) in each permutation $\sigma$
\cite{Renema2018}. When this is done, permanents of positive matrices
arise, which can be approximated efficiently \cite{Jerrum2001}. Grouping
terms by the size of the derangements $j$ (i.e. the number of elements
not corresponding to fixed points) and substituting equation (2) into
equation (1), we have:
\begin{eqnarray}
P & = & {n\choose m}^{-1}\sum_{j=0}^{m}\sum_{\tau}\sum_{\sigma_{j}(\tau)}\sum_{\rho}\mathrm{Perm}(M_{\tau,\rho}\circ M_{\sigma_{j}^{p}(\tau),\rho}^{*})\times\nonumber \\
 &  & \ \ \ \ \ \ \mathrm{\ \ \ \!\ \ \ \ \ \ \ \ \ \ \ \!\ \ \ \ \ \ \ \ \ \ Perm}(|M_{\sigma_{j}^{u}(\tau),\bar{\rho}}|^{2}),
\end{eqnarray}
where we have made use of their independence to exchange the outer
two sums, and where we have used Laplace expansion for permanents to split the matrix permanent into two parts, corresponding to quantum and classical particles, respectively. The notation $\sigma_{j}(\tau)$ denotes a permutation
with $m-j$ fixed points which is constructed from the elements of
$\tau$, $\sigma_{j}^{p}$ is the permuted part of such a permutation
$\sigma_{j},$ $\sigma_{j}^{u}$ is the unpermuted part, $\rho$ is
a $j$-combination of $m$, and $\bar{\rho}$ is the complement of
that combination. It should be noted that since the size of $\sigma_{j}^{p}$
grows with $j$, the sum over $j$ serves to group terms by computational
cost, from easiest to hardest. Simultaneously,
one can interpret the $j$-th term as containing all interference
processes involving precisely $j$ bosons \cite{Rohde2015}.


We show that in equation (3), terms with large $j$ carry
less weight and can therefore be neglected beyond some threshold value
$k$, which depends on the losses and the desired accuracy of the
approximation. Since the $j$-th term represents quantum interference of $j$ bosons, this amounts to showing that boson sampling with losses can be understood as boson sampling of some $k<m$ number of bosons. Therefore, $k$ defines the maximum size of a boson sampler which can be constructed at a given level of losses.

We can formalize this idea by computing the expected value of the
$L_{1}$-distance between our approximation $P'$ and the exact distribution  
$P$, i.e., ${d(P,P')= \sum_q  | P(q)-P'(q)|},$ 
where the sum runs over all collision-free outcomes $q$. In the Supplemental Material, we show that at a fixed ratio $\eta = m/n$, the expected value of $d(P,P')$ over the ensemble of Haar random unitaries is upper bounded by the expected value of $d(P,P')$ as $n$ and $m$ go to infinity while keeping their ratio constant. In that case, the expected value is given by: 
\begin{eqnarray}
\mathbb{E}_{U}\left[d(P,P')\right] \leq \sqrt{\eta^{k+1}/(1-\eta)}.
\end{eqnarray}
We note that since we can only compute the expectation value of the
distance, our algorithm will fail for some fraction of unitaries. 
However, using a standard Markov inequality (see Supplemental Material) one can 
bound the probability that $d(P,P')$ does not satisfy a given bound. 
If $\mathbb{E}_{U}\left[d(P,P')\right] \leq\epsilon$ one can shown that
$P\left[d(P,P')>\epsilon/\delta\right]$ is upper-bounded by $\delta$. We note that numerical simulations suggest (see Supplemental Material) that the scaling in $\delta$ is actually much better than indicated by this bound.

To use our results for sampling, we use our approximation as the input for a
Metropolis sampler that samples efficiently from our approximate
probability distribution, which results in a
classical simulator of boson sampling with losses. We state the algorithm in full: first, given a value of probability
of failure $\delta$, error tolerance $\epsilon$, $n$ and $m$, compute
the maximum boson sampler size $k$ using equation
(5). Second, randomly choose a set of candidate output
modes. Third, compute the approximate output probability
by evaluating equation (3) up to the $k$-th term.
Fourth, use this probability to compute the acceptance
ratio of a Metropolis sampler. Repeat steps 2-4
to generate more samples. In order to compute this approximation,
we need evaluate to evaluate a polynomial number of permanents of size up to
k \cite{Renema2018}.

Solving equation (4) for the maximum boson sampler size $k$ gives
the number of bosons beyond which our algorithm becomes efficient,
given a user-defined probability of failure $\delta$,
error tolerance $\epsilon$ and the value of $\eta$ corresponding
to the experimental setup, which scales as
\begin{eqnarray}
k\approx 2\frac{\log \epsilon+\log \delta + \log (1 - \eta)}{\log \eta},
\end{eqnarray}
which shows that the algorithm is efficient in $\epsilon$ and $\delta$. The fact that $k \rightarrow \infty$ when $\eta \rightarrow 1$, represents the fact that an ideal boson sampler cannot be simulated with our algorithm efficiently, as expected.


We note that equation 5 produces a finite value of $k$ for any level of imperfections $\eta < 1$. Therefore, for any level of imperfections, boson sampling can be appoximated as quantum interference of $k < n$ particles and classical interference of the remaining $k-n$ particles. This result shows that demonstrations of a quantum advantage with boson sampling will be non-scalable for any level of imperfections.

However, in practice, one is interested in performing boson sampling at some finite $n$, which should be larger than what can be simulated with a classical supercomputer. Current estimates put this around 50 photons \cite{Wusupercomputer}. Using these assumptions, the algorithm presented above 
can be used to rule out a quantum advantage in certain
areas of the parameter space. Figure 2 shows the restrictions which
our algorithm places on losses: we show parametric plots solving
equation (5) for $k$ versus $\eta$, for $\mathbb{E}(d)=0.1,$ $0.01$ and
$0.001$. 
We find that in order to have 50-boson
interference while maintaining an average $L_{1}$-distance of $0.1$, a transmission of $\eta>0.88$ is necessary. 
If a higher accuracy of the classical algorithm is required, this results only in a polynomial increase in $k$.

\begin{figure}
\includegraphics[width=8cm]{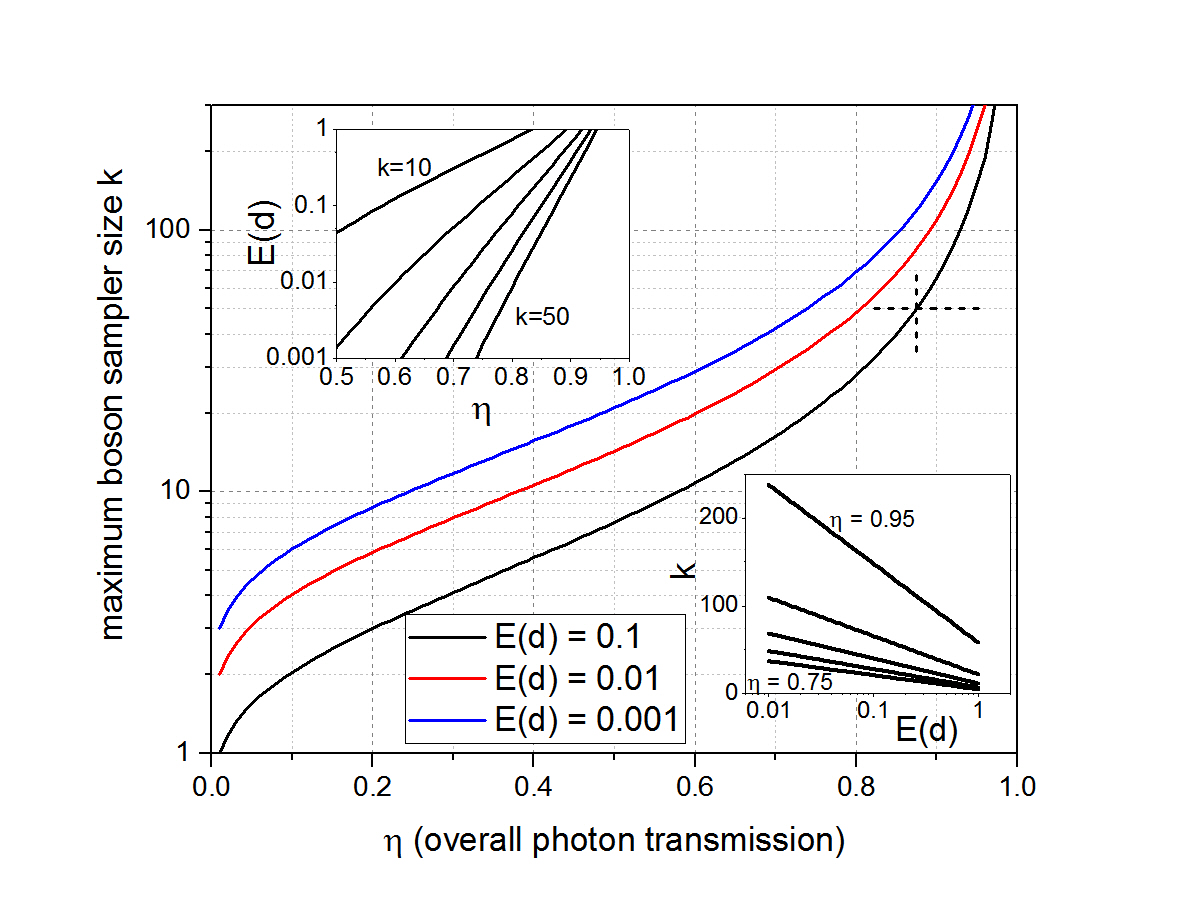}\caption{Maximum number of bosons $k$ which can be interfered before our
classical algorithm becomes efficient, as a function of the overall
boson transmission $\eta,$ including losses in the sources, interferometer
and detectors. The horizontal dashed line indicates the point $n=50,$
which is conventionally taken as the limit of a quantum advantage.
The vertical dashed line indicates the corresponding requirement of
$\eta>0.88$ for $\mathbb{E}(d)=0.1$. \emph{Left inset}: dependence of $\mathbb{E}(d)$
on $\eta$ for $k=\{10,20,30,40,50\}.$ \emph{Right inset}: dependence
of $k$ on $\mathbb{E}(d)$ for $\eta=\{0.75,0.8,0.85,0.9,0.95\}.$ }
\end{figure}



Losses are not the source of noise that a boson sampling device may suffer from.
Experimental partial distinguishability of bosons, which are in principle completely
indistinguishable particles, can have an important impact on the quality of an experiment. Following the treatment of \cite{Renema2018}, our algorithm leads
to a very simple treatment of both sources of noise.
We re-introduce a finite level of boson distinguishability,
as the overlap $x=\langle\psi_{i}|\psi_{j}\rangle$ for $i\neq j$ of the internal (i.e. non-spatial) parts of the wave function of the photons,
where $|\psi_{i}\rangle$ is the internal part of the wave function of the $i$-th boson.
As derived in the Supplementary Material, equation (4) holds, but with $\alpha=\eta x^{2}$ taking the place of $\eta.$ 
Regardless of the specific combination of $\eta$ and $x$ used to achieve it,
our algorithm can approximate experiments with equal $\alpha$ equally
well. Its value may therefore be taken as a figure of merit of the
ability of an experiment to interfere large numbers of bosons.

\begin{figure}
\includegraphics[width=8cm]{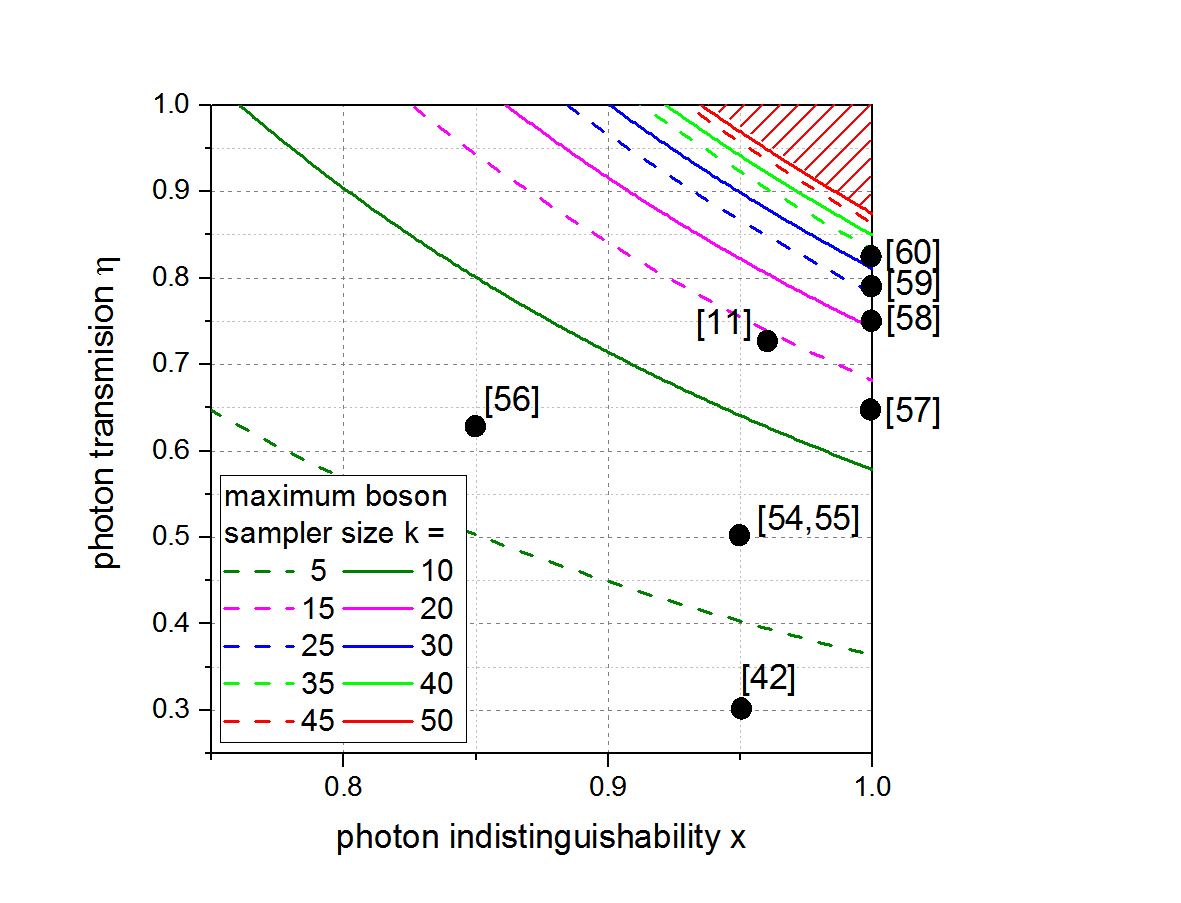}

\caption{Interchange between distinguishability parameter $x$ and the transmission
probability $\eta$, for $\mathbb{E}(d)=0.1$. The lines in this plot are contours
of constant $\alpha=\eta x^{2},$ for $\alpha$ corresponding to the
number of photons indicated in the legend. The red shaded area in
the top right is the region of the parameter space where our algorithm
cannot rule out a quantum advantage demonstration. The black points
indicate the values of $\eta$ and $x$ of various photon sources
reported in the literature.}
\end{figure}

\hyphenation{post-selection}



Figure 3 shows the tradeoff between boson distinguishability and
loss. The curves are plots of $\eta=\alpha/x^{2}$,
where the $\alpha$ correspond to different values of the maximum
boson sampler size $k,$ as indicated in the legend. The black dots
incidate various photon sources reported in experiments \cite{Wang2017,Sparrowquantum,Quandela,Gazzano2013,Zhong2018,Dresdendot,Shalm2015,Giustina2015,Slussarenko2017}.
The plot was generated for $\mathbb{E}(d) = 0.1$. This plot therefore demarcates
areas of the parameter space where interference of a given number
of photons cannot be simulated at that level of accuracy.

We observe that today, not even the best possible experiments meet the requirements for
a scalable demonstration of quantum advantage: considering the best
interferometers (99\% transmission) \cite{Wang2018}, the best detectors
(93\% efficiency) \cite{Marsili2013} and the best photon sources, we
arrive at $\alpha=0.755$, which implies that any scalable boson sampling
experiments with more than $21$ photons will be simulable
with our algorithm at the 10\% accuracy level.

In generating Figure 3, the relevant quantity for heralded photon sources (such as those operating 
on parametric down conversion) is the heralding efficiency, i.e. the probability that a photon enters the experiment given
that the source signals that it has produced a photon. The heralding efficiency on parametric down conversion sources can be as high as 82\% \cite{Slussarenko2017}, but the overall photon generation probability is typically a few percent, to avoid multiphoton generation. While for boson sampling, this does not affect the complexity
\cite{Lund2014}, further photonic quantum technologies will most likely require photon sources with a high absolute photon generation probability.

Finally, we consider proposals involving postselection on those cases
where (almost) all bosons make it through, as is usually done in
experiments. Wang \emph{et al. } recently proposed such an experiment,
generating $50+p$ photons and detecting $50$ \cite{Wang2018}. Our
results show than in such a case (see Supplemental Material), $p\leq3$ is required in order not
to be simulable by our algorithm at the 10\% approximation level.

In real experiments, $m$ fluctuates
according to a binomial distribution with mean $\eta n$
and variance ${\eta(1-\eta)n}$.
One can always efficiently simulate these fluctuations if one has
an algorithm for a fixed pair $m$ and $n$, since there are approximately $\sqrt{m}$
possible outcomes which occur with high probability, and these are
clustered around $m=\eta n$. As shown in the Supplementary Material, 
such a postselection only adds a small correction to equation (5) and
a prefactor to our bound, which vanishes in the limit
of large $n$. Therefore, the classical algorithm starts by estimating a
$\tilde{k}$, given a value of
probability of failure $\delta$,
error tolerance $\epsilon$, and $n$. To generate samples, we simply modify our algorithm to sample over $m$ according to the binomial distribution.


Recently, an estimate of the loss which compromises the demonstration
a quantum advantage was given by Neville \emph{et al. }\cite{Neville2017}\emph{.
}Our result improves on this result in several ways. First, we demonstrate
how losses induce a transition in the scaling of the runtime of classical simulation of a boson sampler from exponential to polynomial, while their result is essentially a runtime
estimate comparing an inefficient classical calculation against an
inefficient experiment. Second, because our algorithm is polynomial in the number of particles,
the bounds that we find are also much more stringent in an absolute
sense. Third, we show that the required transmission is a monotonically
increasing function of the number of bosons which is coherently interfered. 
Whereas \cite{Neville2017} was only able to show that lossy boson sampling is at most
as hard as regular boson sampling, we show that it is in fact much easier.

Future improvement to this work can by made by generalizing our results
to non-uniform losses and more general disitinguishability
models and replacing the Metropolis sampler
by a direct sampling algorithm, such as the one proposed
by Clifford and Clifford \cite{Clifford} for exact boson sampling. An
adaptation of this result to Gaussian boson sampling, an
alternative approach for quantum supremacy that has application
as a subroutine in a classical-quantum hybrid
algorithm for the calculation of the vibronic spectra of
molecules and finding dense subgraphs, would also be an
interesting future research direction.

In conclusion, we have found an efficient classical simulation algorithm for a quantum advantage proposal, namely boson sampling. This algorithm uses the natural sources of noise present in the system, i.e. linear photon loss and particle distinguishability. It predicts an upper bound to the size of a boson sampler that can be constructed with noisy components. Remarkably, that upper bound only depends on a scale-invariate parameter combining losses and indistinguishability, which allows us to demarcate areas of the parameter space where a quantum advantage is impossible. By evaluating this bound for state-of-the-art photonic components,  we find that with current technology, boson sampling cannot demonstrate a quantum advantage. 

\subsection*{Acknowledgements}

We thank Pepijn Pinkse for critical reading of the manuscript. We thank Dirk Englund for discussions. J.J.R. acknowledges support from
NWO Vici through Pepijn Pinkse. V.S. acknowledges CNPq of Brazil grant
304129/2015-1 and FAPESP grant 2018/24664-9 . R.G.P. Acknowledges the support of F.R.S.- FNRS and the Fondation Wiener Anspach.

\subsection*{Author contributions}
J.J.R. conceived the work. V.S. and R.G.P. contributed to the proof.
All authors contributed to the writing of the manuscript.

\subsection*{Competing interests}
The authors declare no competing interests.

\end{document}


\providecommand{\tabularnewline}{\\}

\makeatother

\newtheorem{theorem}{Theorem}

\title{Supplemental material for: Classical simulability of noisy boson sampling}

\maketitle

\section{The boson sampling problem}

The boson sampling problem can be stated as follows: given $n$ photons incident on a network of coupled interferometers containing $N$ modes, provide a sample from the photon distribution at the outcome (i.e. a sample of the detection events observed). This problem is of note because it is strongly believed to be efficiently solvable on a quantum machine which implements the problem directly, while being intractable for a classical computer \cite{Aaronson2011}. 

Calling the unitary transformation of the interferometer $U$, labeling the input modes $\tau$ and the output modes $q$, we can describe the probability of a given outcome in an ideal boson sampler (i.e. one without imperfections) as 

\begin{equation}
P(q|\tau) = |\mathrm{Perm}(M)|^2,
\end{equation}
where $\mathrm{Perm}$ is the matrix permanent function, and $M$ is a matrix constructed by taking the elements of $U$ which connect $\tau$ and $q$, i.e. select rows from $U$ according to the indices of $\tau$ and columns according to $q$. We will assume throughout that we are working in the collision-free regime, where the probability of two photon emerging from the same mode can be neglected. 

If we add imperfections to this model, equation (1) is modified. For losses, we must consider all ${n \choose m}$ ways in which $n$ input photons could give rise to $m$ detection events. Hence we have:

\begin{equation}
P(q) = {n \choose m}^{-1} \sum_\tau |\mathrm{Perm}(M)^2| = {n \choose m}^{-1} \sum_\tau P(q|\tau),
\end{equation}
where the sum runs over all possible $\tau$, and the coefficient of normalization is chosen such that the distribution postselected on the arrival of precisely $m$ photons is normalized. 

Finally, if we add photon distinguishability (nonunit wave function overlap), the probability of observing a given event is modified:

\begin{equation}
P(q|\tau) = \sum_\sigma (\prod_i S_{i,\sigma_i}) \mathrm{Perm}(M_{\tau,1} \circ M^*_{\sigma(\tau),1}),
\end{equation}
where $S_{i,j}$ is the wave function overlap between the $i$-th and $j$-th photons, $\sigma$ is a permutation, and the sum runs over all permutations. Equations 2 and 3 can be combined to take into account the effect of loss and distinguishability simultaneously.

\section{Proof sketch}

We provide a sketch of the proof of equation 5 of the main text. A full derivation can be found in sections VII-X below. Our proof can be divided in three parts. First, we reduce the problem of approximating the distribution of a boson sampler with imperfections to the problem of finding an approximation for a single outcome of the boson sampler with a bounded expectation value of the relative error on the approximation. Then, we construct such an algorithm and compute its expected error. Finally, we show how to use this algorithm for sampling.

The first step begins with the observation that, as a function of the size of the sampler, we wish to maintain a constant distance $\epsilon$ between the probability distribution given by our approximation and the exact probability distribution, i.e. $d = \sum_q |P'(q) - P(q)| < \epsilon$, where $q$ denotes a possible output (i.e. pattern of detection events) of the boson sampler, $P'(q)$ is the probability of $q$ predicted by our approximation, and the sum runs over all outputs. 

However, since the number of outputs increases as the size of the boson sampler increases (either in the number of particles or in the number of modes), it is insufficient for each term within this sum to have a constant error with increasing boson sampler size: we must find an algorithm where the error decreases so as to compensate for the number of increasing terms in the sum. This means that in order for any approximate algorithm to be useful, the error scaling of the that algorithm must be inversely propotional with the number of outcomes of the boson sampler.

Next, we show that it is sufficient to merely bound the expected value of this error over the set of all unitary matrices. Since the trace distance is bounded from below at 0, the Markov inequality applies, which says that for a non-negative probability distribution, no more than $1/k$ of a probability distribution $d$ can lie beyond $kE(d)$. Use of the Markov inequality allows us to average over the set of unitary matrices, which is convenient because in the limits in which boson sampling operates, submatrices of the unitary matrix $U$ consist of i.i.d. Gaussians, which are very convenient for computing statistical quantities such as expected values and covariances. The price we pay is that the Markov bound is probabilistic: when applying the approximation algorithm to some unitary matrices, their error will lie beyond the expected value. However, we can bring this failure probability down by considering a looser bound, e.g. not considering $E(d)$ but rather $2E(d)$. Finally, we will see that numerics suggests that the Markov bound grossly overestimates the failure probability our our algorithm. 

In the second step, we demonstrate an algorithm which meets the requirements derived in the first step. This algorithm is based on the expansion of $n$-photon interference into terms corresponding to $0, 2, 3, \dots m$-photon interference terms, where $m$ is the number of photons which emerge at the output of the boson sampling machine. We show that we can write the probability of a given outcome as a polynomial: 

\begin{equation}
P(q) = \sum^m_{j=0} c_j x^j,
\end{equation}
where $x$ is the photon distinguishability (we simplify the problem by assuming all photons are equally distinguishable) and $c_j$ is a coefficient, which is a function of the unitary matrix $U$, as well as the degree of loss $\eta = m/n$. Expressions for computing $c_j$ are given in the supplemental material.

The central intuition of our result is that in this expression, any imperfection degrades the higher-order terms of the polynomial the strongest. This happens either directly, since finite distinguishability corresponds to $x < 1$, or through the coefficients $c_j$ in the case of photon loss. 

In Figure 1, we show this effect. As an illustration, we numerically compute the variance of $c_j x^j$ for 3 different scenarios: a lossless boson sampler with 6 photons, a boson sampler with 8 input photons and 6 output photons (i.e.75\% transmission) or a boson sampler with 0.75 Hong-Ou-Mandel visbility (which corresponds to $x^2 = 0.75)$. To generate this figure, we simulated 10000 boson sampling experiments under these three sets of conditions. We have normalized each graph such that $ \mathrm{Var}(c_0) = 1$, to facilitate comparison between the curves. The decrease in size between the terms for the perfect boson sampler for increasing $j$ is due to variations in the number and expected value of terms which make up $c_j$, and does not affect our results.

\begin{figure}
\includegraphics[width = 9 cm]{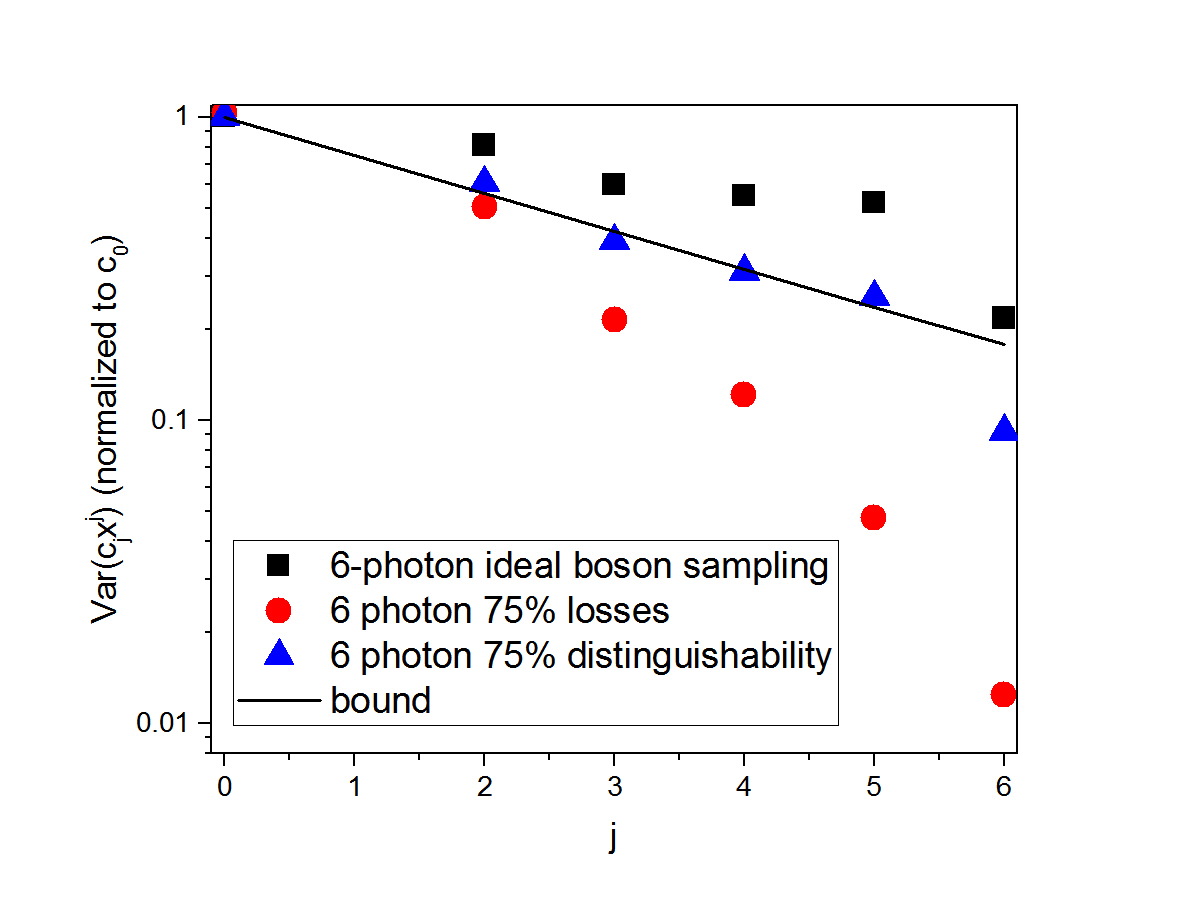}
\caption{An illustration of the central intuition of our results: the weight of higher photon number interference is suppressed by either loss or distinguishability. We performed numerical simulations of 10000 boson sampling experiments, for three cases: the black squares represent an ideal 6-photon experiment, the red circles represent an 8-photon experiment where 6 photons are transmitted, and the blue dots represent an experiment with 6 photons with 75\% Hong-Ou-Mandel visibility. The line is the bound which we use in our calculations.}
\end{figure}

As can be seen from Figure 1, inperfections exponentially dampen the higher-order terms in this expansion. This suggests neglecting the higher order terms by truncating the sum at some index $k$, thereby producing an approximation of the form $P'(q) = \sum^k_{j=0} c_j x^j$. In order to compute the effect of such a trunctation, we can bound the variance of $c_j$ with a bound of the form $\mathrm{Var}(c_j) < (\eta x^2)^j$, where $\eta$ is the transmission of the boson sampler. We have indicated this bound as a solid line in Figure 1. Note that this bound takes into account losses and distinguishability simultaneously, which is a powerful feature of our algorithm.

Using this bound, we can now compute the error of our approximation for computing a single output probability. We find that it has the form required to result in a constant error on the expected value of the complete distribution. We find that the expected value of the error $d = \sum_q |P'(q) - P(q)|$ is bounded by $E(d) < \sqrt{\alpha^{k+1} / (1-\alpha)}$, where $\alpha = \eta x^2$. This means that if we apply a truncation at $k$-photon interference, we can expect the resulting distribution to be typically no further away from the ideal one than given by this expression. 

In the last major step of our proof, we convert this approximation into a sampling algorithm. We make use of the results by \cite{Neville2017}. In that work, the authors showed that using a Markov-Chain Monte Carlo (MCMC) technique, one can obtain samples from a boson sampler at the cost of computing a fixed number of permanents. 

We present a simplified version of the MCMC from \cite{Neville2017}. MCMCs work by recording a position in the parameter space of outputs $q$. In every iteration, some new candidate output is chosen, and the sampler moves to that position with probability $\mathrm{min}(1,P(q_{new})/P(q_{old}))$, i.e. it always moves to the candidate position if it has higher probability, and otherwise it moves with some probability. At the end of each iteration, the present position of the sampler is output as a sample. It is known that such a sampler samples from the probability distribution $P(q)$. 

In our case, our approximate sampler samples from the approximate probability distribution $P'(q)$ according to the method outlined above. There are now two complications. First, $P'(q)$ may occasionally be negative, or even greater than 1, since it is only an approximation of a probability distribution. This is easily taken care of by constraining $0 \leq P'(q) \leq 1$: since this same condition holds for $P$, this will only decrease the distance between the two. The second condition is that each sample must be generated efficiently, i.e. in polynomial time. To show this is the case, we need to look into the computational cost of the $c_j$. We can show that the computational cost of the truncation scales as $m^{2k}$. To achieve this number, it is required that we do not merely sample over the potential output modes $q$, but also over the input modes, i.e. over all possible sets of photons that could possibly have given rise to a given output configuration.

Finally, we must also sample over the distribution of possible numbers of photons which make it through the boson sampler in the first place. This can be done by running $n$ classical sampling algorithms of the form described above in parallel, and drawing a sample from one of them according to the binomial distribution describing the probability of observing a certain number of photons. In the supplemental material, we show that these fluctuations in photon number do not substantially affect our bound, which was derived for a fixed number of output photons. This concludes the proof.

\section{Boson sampling with constant loss}

Wang \textit{et al} recently proposed an experiment where 50 photon boson sampling would be attained with realistic improvements on current technology, by allowing for the loss of a few bosons \cite{Wang2018}. In such an experiment, 50 + $p$ photons are sent into the interferometer, and postselection on 50 detection events is performed. For elementary reasons of combinatorics, given a fixed repetition rate of the experiment, the higher $p$, the higher the rate with which samples are obtained. 

Using our results, we can bound $p$, and therefore bound the production rate of 50-photon samples in an experiment in the spirit of \cite{Wang2018}. In the case of postselected experiments, $\eta$ is given by the ratio of $m$ and $n$ that is enforced by the postselection, rather than by the transmission of the experiment. Furthermore, we will assume that the distinguishability of the photons used in such an experiment remains at $x^2 = 0.939$. Therefore, equation (5) of the main text reads:

\begin{equation}
\frac{x^{2(k+1)}\left(\frac{n}{n+p}\right)^{k+1}}{1 - (x^2 \frac{n}{n+p})}\geq \mathbb{E}(d)^2.
\end{equation}

Solving this numerically for $p$, assuming $n=50$, $k=49$, $x^2 = 0.939$ and $\mathbb{E}(d) = 0.1$ yields $p \leq 3.665$, meaning that at most 3 photons can be lost in this experiment without becoming classically simulable according to our criteria.

Interestingly, due to the interplay between photon losses and distinguishability in our algorithm, creating photons with perfect distinguishability would actually increase the number of photons which can be lost before becoming classically simulable. If we assume perfect photons, i.e. $x = 1$, then lossy boson sampling with up to 7 photons lost would be permissible under our algorithm, which would correspond with sampling rates in excess of 1/s \cite{Wang2018}. 

\section{Numerical results}

\begin{figure}[ht]
\includegraphics{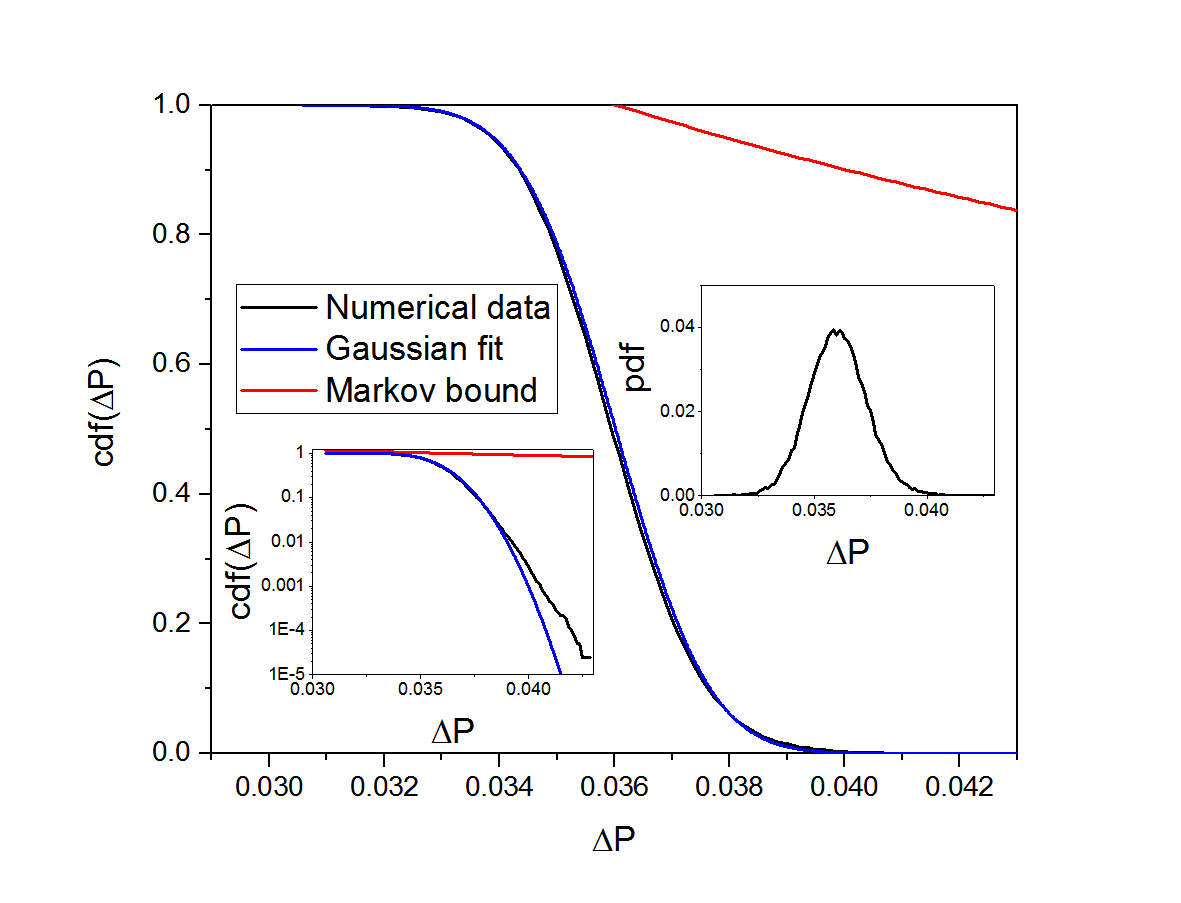}

\caption{Cummulative probability distribution of $d=\sum\limits_q |P_m(q)-P^\prime_m(q)|$ computed for 40000 unitary matrices, with $N = 15$, $n = 5$, $m = 3$, $x = 1$, $k = 1$. The black line shows our numerical data. The red line shows the Markov bound, and the blue line shows a Gaussian fit to our numerical data. This result demonstrates that the probability distribution of $d$ over the set of $U$ is much more clustered around its expectation value than the Markov bound would suggest. Left inset: the cummulative probability distribtion on a log scale. This inset highlights the excess probability above $E(d)$ compared to a Gaussian distribution. Right inset: the (numerically evaluated) probability density function of $d$.}

\end{figure}

We use a Markov inequality to upper bound the fraction of unitaries for which the distance between the set of output probabilities and our approximation exceeds the expected value computed in equation 4 by some margin, i.e. those which have a value of $d = \sum_q |P(q) - P^\prime (q)| > \alpha \mathbb{E}(d)$. However, numerical evidence suggests that this bound is far from optimal. Figure 2 shows a numerical simulation, where we computed $d$ for 40000 Haar-random unitaries. We set $N = 15$, $n = 5$, $m = 3$, $x = 1$, $k =1$, but the result is representative for other choices of the simulation parameters. This result shows that our Markov bound strongly overestimates the number of events which exceed $\mathbb{E}(d)$ by a large margin. 

We can provide the following explanation for this result: most output configurations $q$ are completely uncorrelated. Therefore, fluctuations on the quantity $d = |P(q)-P^\prime(q)|$ 
have the tendency to average out when summing over all possible output
configurations. Our results suggest that convergence to a Gaussian is not complete: we observe a slight (but significant) increase of events which have larger $d$ than predicted by our Gaussian fit, which is shown in Figure 1.  We leave the study of this  phenomenon to future work. 

\section{Table for Figure 3}

For completeness, we present the data used to generate Figure 3 of the main text in tabular form. 

\begin{table}[htb]
\begin{tabular}{|c|c|c|c|c|c|}
\hline 
Reference & Type & $\eta$ & $x^{2}$ & $\alpha$ & k\tabularnewline
\hline 
\hline 
\cite{Wang2017} & QD & 0.3 & 0.94 & 0.282 & 3\tabularnewline
\hline 
\cite{Sparrowquantum,Quandela} & QD & 0.5 & 0.95 & 0.475 & 7\tabularnewline
\hline 
\cite{Gazzano2013} & QD & 0.62 & 0.85 & 0.533 & 8\tabularnewline
\hline
\cite{Dresdendot} & QD & 0.65 & 0.998 & 0.65 & 13 \tabularnewline
\hline
\cite{Zhong2018} & SPDC & 0.73 & 0.96 & 0.67 & 16 \tabularnewline
\hline 
\cite{Shalm2015} & SPDC & 0.75 & $\approx1$ & 0.75 & 20\tabularnewline
\hline 
\cite{Giustina2015} & SPDC & 0.79 & $\approx1$ & 0.79 & 26\tabularnewline
\hline 
\cite{Slussarenko2017} & SPDC & 0.82 & $\approx1$ & 0.82 & 31\tabularnewline
\hline 
\end{tabular}

\caption{The experimental data used to generate
Figure 3 from the main text, given in tabular form:  The figure of merit $\alpha$ for various photon sources,
as well as the maximally achievable number of photons that can be effectively
interferered.}
\end{table}

\section{Code availability}
An implementation of our algorithm can be found online at https://github.com/jrenema/BosonSampling

\section{Full derivation}

In the following sections, we provide a derivation of equation 4 of the main text. Our proof proceeds as follows. First, we demonstrate that in order to approximate the whole probability distribution of outcomes from a boson sampler, it suffices to be able to approximate the probability distribution of a single outcome to within constant relative error. Then, we demonstrate that it suffices to bound merely the expected value of the error on the approximation. Next, we give an algorithm for approximating a single outcome with a bounded expected value of the relative error. Having done that, we can state our results on classical simulability of boson sampling in the form of 2 theorems.

\section{From distribution to a single outcome} 

We begin by recalling the problem at hand and the main assumptions. We are interested in classically approximating a boson sampler with the imperfections of partial photon distinguishability and non-unit transmission. We assume all photons are in internal states $|\phi_i\rangle$, $i=1,\ldots, n$, with a uniform overlap $x  \equiv \langle \phi_i|\phi_j\rangle$ for $j\ne i$, and hence the distinguishability is parameterized by a single real number $0\le x\le 1$. 
We assume that the loss of photons is path independent, and that it can therefore be assumed to occur at the input of an optical network which implements the unitary $U$. Under that assumption, the probability for $m$ single photons to emerge from the system given that $n$ were input is given by $\left({n \atop m}\right)\eta^m(1-\eta)^{n-m}$, where $0<\eta<1$ is the transmission coefficient. Furthermore, we shall assume that the number of modes $N$ is chosen such that the probability of two photons emerging from the same mode may be neglected (which is the case when $N \gg n^2$). We shall therefore only consider outputs where all photons emerge from distinct modes.

Without loss of generality, we may assume that the photons are incident on the first $n$ modes. In order to compute the probability of observing a particular output configuration $q = (l_1,\ldots,l_m)$, we must classically sum over all possible ways which any subset $\tau = (k_1,\ldots,k_m)$, $k_i< k_{i+1}$ of $m$ photons out of $n$ could have given rise to this configuration. Calling the probability of $\tau$ causing $q$ $P(\tau | q)$, we have:  

\begin{equation} 
P = \left({n \atop m}\right)\eta^m(1-\eta)^{n-m}P_m(q),
\end{equation}
with:
\begin{equation}
P_m(q) =\left({n \atop m}\right)^{-1}\sum_{\tau}P(\tau | q),
 \label{V1}
\end{equation}
where $P_m(q)$ is chosen such that the probablity distribution for a fixed $m$ is normalized to $\sum_q P_m(q) = 1$. $P(\tau | q)$ is a function of the matrix $M$, which is a submatrix of $U$ consisting of those elements containing the probability amplitudes connecting $\tau$ and $q$, as well as the distinguishability of the bosons. Recall that the boson sampling complexity proof requires a randomly chosen $N$-port unitary network $U$  (in the Haar measure over the unitary group). In the limit that $N \gg n^6$, the matrix elements of $M$ can be well approximated by i.i.d. Gaussians \cite{Aaronson2011}, and we shall assume this is the case. 

We will estimate the trace (i.e. $L_1$) distance $d$ between the probability distribution of boson sampling with a lossy network ($P$) and another probability distribution ($P^\prime$), which we will construct as an approximation of $P$, and which we will construct such that it can be efficiently sampled from with classical resources. We consider two cases. First, we concentrate on the post-selected case of a  \textit{fixed}  number of detected photons $m$, which is relevant to some recent experimental proposals (e.g., Ref. \cite{Wang2018}).    In this case we have 
\begin{equation} 
\label{V2}
d_m =  \sum_q |P_m(q) - P^\prime_m(q)|,
\end{equation}
where  $q$ runs over all distinct combinations of the output ports, $q = (l_1,\ldots,l_m)$, $l_i< l_{i+1}$. Our main interest  is the  trace distance between the full distributions, i.e., with a \textit{random} number of detected photons $m$:
\begin{equation} 
\label{V3}
d =  \sum_{m=0}^n \left({n \atop m}\right) \eta^m(1-\eta)^{n-m} d_m.
  \end{equation}

To construct an approximation of $P_m(q)$  that one can simulate classically,  we expand the boson sampling distribution over  the  number of effectively  interfering  photons $0\le j\le m$  (for more details on the orders of interference, see the recent Ref. \cite{Valery2018})    and choose $P^\prime_m(q)$ to be the sum of the obtained terms up to $j=k$ for some  $k$ that guarantees a polynomial algorithm to approximate the lossy boson sampling to a given error in the trace distance in Eq. (\ref{V3}). Indeed, since we  take    $k$ out of $n$ input photons (i.e., only $k$ quantum phartially distinguishable bosons) and the rest treat as the  classical particles (i.e., bosons in internal states orthogonal  to each other and to the rest of the bosons in the system) such an algorithm is possible (see below). 

Our method is to bound  the terms in the trace distance (equation \ref{V2}), i.e.,  the difference  $|P_m(q)-P^\prime_m(q)|$ for  $q = (l_1,\ldots,l_m)$, $l_i< l_{i+1}$,  for a randomly chosen network, i.e. by considering an average in the  Haar measure over the group of unitary networks $U$. We begin by noting that the probability of a single outcome in the Haar measure is given by $\langle P_m(q)\rangle  =  \langle P(\tau|q)\rangle = \frac{m!}{N^m} \equiv \langle P_m\rangle$ for any state of partial distinguishability of photons (see, for instance, Ref. \cite{Valery2014}), which is also  the number of terms in Eq. (\ref{V2}). A simple counting argument gives an informal reason why this is so: if the probability of a collisionless outcome is negligible, the total number of outcomes is given by ${N \choose m}$, and by symmetry, these are all equally likely. Therefore, the average probability of any one outcome is given by ${N \choose m}^{-1} \approx \frac{m!}{N^m}.$

Because of this, a   bound on the variance in the form 
\begin{equation}
\label{V4A}
\mathrm{Var}(P_m(q) - P^\prime_m(q)) \le \langle P_m \rangle^2 \epsilon^2
\end{equation}
 implies also  a bound on the average trace distance  error of Eq. (\ref{V2}): $\langle d_m\rangle \le \epsilon$. Indeed,  for a real random variable $X= P_m(q) - P^\prime_m(q)$  of  zero average $\langle X\rangle =0$ we have $\langle |X|\rangle \le \sqrt{\mathrm{Var}(X)}$.  Hence  $\langle |P_m(q) - P^\prime_m(q)|\rangle \le     \epsilon \langle P_m \rangle $ implies   $\langle d_m\rangle \le \epsilon$ by the fact that the average $\langle P_m \rangle$ is the inverse of the number of terms in Eq. (\ref{V2}). A similar bound on the trace distance between the full distributions with random $m$,  Eq. (\ref{V3}),  can be  derived as well (see below). 

Here we note that by considering an average in the  Haar measure over  unitary networks $U$  we allow a small probability of failure of our approach, i.e., for some small fraction of the networks $U$.   By the standard means, e.g.  using   Markov's  inequality in probability, we obtain  for any $a>1$
\begin{eqnarray}
 \label{V4}
&& Prob\left\{ |P_m(q) - P^\prime_m(q)| > a\epsilon \langle P_m \rangle\right\}\le \frac{   \mathrm{Var}(P_m(q) - P^\prime_m(q))}{(a\epsilon \langle P_m\rangle)^2}\le \left(\frac{1}{a}\right)^2,\nonumber\\
&& Prob\left\{ d_m > a\epsilon  \right\}\le \frac{\langle d_m\rangle }{a\epsilon} \le \frac{1}{a}
\end{eqnarray}
where we have used Eq. (\ref{V4A}).  

\section{Approximating a single probability}

From Section 2, we have seen that finding the variance $\mathrm{Var}(P_m(q) - P^\prime_m(q)) $ is the key to our approach. To estimate the variance on the r.h.s. of Eq. (\ref{V4}), our strategy is to expand the probability $P(\tau| q)$  of Eq. (\ref{V1})  into the terms that describe the interferences  of $j$ photons (i.e., term $j$ contains  the permutations    that have exactly $m-j$ fixed points, i.e., non-interfering photons). 

We begin with the following expression for $P(\tau | q)$: 

\begin{equation}
P(\tau | q) = \sum_\sigma \left( \prod_i^m S_{i,\sigma(i)} \right) \mathrm{Perm} \left(M_{\tau} \circ M^{*}_{\sigma(\tau),1}\right),
\label{V4B}
\end{equation}
where $S_{ij} = x + \delta_{ij}(1-x)$, $\mathrm{Perm}(A) = \sum_{\sigma} \prod_i A_{i,\sigma(i)}$ is the matrix permanent function, $M_{\tau}$ is the sub-matrix corresponding to input $\tau$ and output $q$, $\sigma(\tau)$ denotes permutation of the elements of $\tau$ according to the permutation $\sigma$, $M_{\sigma,1}$ denotes permutation of the rows of $M$ according to $\sigma$ and the columns according to the identity, $*$ denotes the complex conjugate, and $\circ$ denotes the elementwise product. In this expression, as in all others in this section, we will suppress the explicit dependence on the output configuration $q$, and take all sums over permutations over all permutations in that set. 

We may group the permutations in the outer sum of equation \ref{V4B} by their number of derangements (unpermuted elements). Denoting the number of fixed points (i.e. unpermuted elements) as $m-j$, we can rewrite equation \ref{V4B} as:

 \begin{equation}
 \label{V5}
P(\tau | q) =\sum_{j=0}^{m}x^j \sum_{\sigma_{j}}\mathrm{Perm}(M_{\tau}\circ M_{\sigma_{j}(\tau),1}^{*}),
\end{equation}
where the notation $\sigma_{j}$  indicates a permutation with  $j$ derangements.  By performing Laplace expansion of the matrix permanents in Eq. (\ref{V5}) on the  rows corresponding to the  fixed $m-j$ 
points, we obtain:
\begin{eqnarray}
\label{V6}
P(\tau | q)& = & \sum_{j=0}^{m}x^j\sum_{\sigma_{j}}\sum_{\rho}\mathrm{Perm}(M_{\tau,\rho} \circ M_{\sigma_{j}^{p}(\tau),\rho}^{*})\mathrm{Perm}(|M_{\sigma_{j}^{u}(\tau),\bar{\rho}}|^{2}),
\end{eqnarray}
where $\sigma_{j}^{p}$ is the set of derangements of  the permutation $\sigma_{j} $ (i.e., its  ``permuted part''), 
$\sigma_{j}^{u}$ is the  set of fixed points, $\rho$ is an $(m-j)$-combination
of $m$, and $\bar{\rho}$ is the complement of that combination.
Using Eq. (\ref{V6}) let us expand the probability in Eq. (\ref{V1}) as follows (again,  omitting the output port  indices $q$, for simplicity):
\begin{eqnarray}
\label{V7}
P_m(q) & = & \left({n\atop m }\right)^{-1}\sum_{j=0}^{m}x^j c_{j},  \\
 & = & \left(\begin{array}{c}
n\\
m
\end{array}\right)^{-1}\sum_{j=0}^{m}x^j\sum_{\tau}\sum_{\sigma_{j}}R_{\tau,\sigma_{j}}, \nonumber \\
 & = & \left(\begin{array}{c}
n\\
m
\end{array}\right)^{-1}\sum_{j=0}^{m}x^j\sum_{\tau}\sum_{\sigma_{j}}\sum_{\rho}S_{\tau,\sigma_{j},\rho},\nonumber \\
 & = & \left(\begin{array}{c}
n\\
m
\end{array}\right)^{-1}\sum_{j=0}^{m}x^j\sum_{\tau}\sum_{\sigma_{j}}\sum_{\rho}\sum_{\xi}T_{\xi,\tau,\sigma_{j}^{u},\rho}V_{\tau,\sigma_{j}^{u},\rho}, \nonumber 
\end{eqnarray}
where $R$ is given by $R = \sum_{\rho}\mathrm{Perm}(M_{\tau,\rho} \circ M_{\sigma_{j}^{p}(\tau),\rho}^{*})\mathrm{Perm}(|M_{\sigma_{j}^{u}(\tau),\bar{\rho}}|^{2})$, $S$ by $S_{\tau,\sigma_{j},\rho}=\mathrm{Perm}(M_{\tau,\rho} \circ M_{\sigma_{j}^{p}(\tau),\rho}^{*})\mathrm{Perm}(|M_{\sigma_{j}^{u}(\tau),\bar{\rho}}|^{2})$,
$T$ by $T_{\xi,\tau,\sigma_{j}^{u},\rho}=\prod_{i}^{m}(M_{\tau,\rho} \circ M_{\sigma_{p}^{u}(\tau)}^{*})_{1,\xi}$
and $V_{\tau,\sigma_{j}^{u},\rho}=\mathrm{Perm}(|M_{\sigma_{j}^{u}(\tau),\bar{\rho}}|^{2}),$
where $\xi$ is a permutation of the $m-j$ elements of $\rho$ and
the sum runs over all such permutations. 

We will now compute the variance of $c_j$.  For this, we will rely on two results from \cite{Renema2018}. The first set of results relates to the covariance of $R$. Two $R_{\tau,\sigma_j}$ terms are only correlated if they share the permuted part of their permutation, i.e. $\mathrm{Cov}(R_{\tau,\sigma},R_{\mu,\nu}) = 0$ unless $\sigma^p = (\nu^p)^{-1}$ \cite{Valery2014,Renema2018}, where $\mathrm{Cov}$ is the covariance, $\mu$ is an $m$-combination of $n$, and $\nu$ is a permutation. Secondly, we will later on need the fairly straightforward computations of the variance of $R$, $S$, $T$ and $V$. 

The reason for this remarkable and crucial fact about covariances is the behaviour of the complex phases in equation \ref{V4B}: if $\sigma^p = (\nu^p)^{-1}$, all phases cancel against each other, and we obtain some finite covariance. If this is not the case, there is a residual phase term which averages out to zero when taking the average over the ensemble of Haar-random unitaries. Note that this condition does not place any restrictions on the phases of the unpermuted parts of $\sigma$ and $\nu$. The reason for this is that the unpermuted parts of $\sigma$ correspond to a permanent of absolute values (cf equation \ref{V6}). From this analysis, it also follows that $R_{\tau,\sigma} = R^*_{\tau,\sigma^{-1}}$. We may therefore, without loss of generality, consider only the real part of $R$ \cite{Shchesnovich2015b}, which we will denote $R$ so as to not make the notation too cumbersome. This concludes the set of results from \cite{Renema2018} that we will use in our derivation.  

From these considerations, it immediately follows that the $c_j$ are uncorrelated, since by construction, they contain terms with derangements of different size, which can therefore never be each other's inverse. Because they are uncorrelated, each $c_j$ contributes independently to the variance used in our bound (equation \ref{V4}), and we can use the expansion into smaller terms given by equation \ref{V7} for a direct computation of these coefficients.  

 The  variance  is given  as
 \begin{equation}
 \label{V8}
\mathrm{Var}(c_{j})=\sum_{\tau,\sigma_{j}}\mathrm{Var}\left(R_{\tau,\sigma_{j}}\right)+\sum_{(\tau,\mu,\tau\neq\mu)}\sum_{(\sigma,\nu,\sigma\neq\nu)}\mathrm{Cov}\left(R_{\tau,\sigma_{j}}, R_{\mu,\nu_{j}}\right),
\end{equation}
where the sums run over all allowed combinations of indices, and where
$\mu$ is an $m$-combination of $n$, and $\nu_{j}$ is a permutation
with $j$ fixed points. 

Our goal then is to understand the behaviour of the (co-)variance of $R_{\tau,\sigma}, R_{\mu,\nu}$, since this allows us to evaluate equation \ref{V8}. For the lossless case, i.e. $n=m$, the condition $\sigma^p = (\nu^p)^{-1}$ enforces that $\sigma = \nu^{-1}$, since in that case, there is only one $\tau = \{1\dots m\}$, and the assignment of the remaining part of the permutation is entirely fixed by the remaining $n-j$ elements of $\tau$. Therefore, in that case, each $R_{\sigma}$ has at most one $R_{\mu}$ that it has any covariance with \cite{Renema2018,Valery2014}. 

However, when we add losses, the picture is complicated somewhat, since we now have freedom to choose $\sigma^u$ and $\nu^u$ by choosing different $\tau$ and $\mu$. Considering a fixed $\tau$ and $\sigma$ in the covariance term of equation \ref{V8}, each $\mu$ which contains the elements of $\nu^p$ now contributes to that covariance term.  These permutations have different fixed points, but identical phase factors, and therefore contribute to the covariance (since they add in phase). 

This is best illustrated with an example. Suppose we set $n=4, m=3$. We therefore have a matrix $M$ of size 4 by 3. Consider two choices of $m$-combinations of $n$ $\tau=\{1,2,3\}$ and $\mu=\{1,2,4\}$. Now consider the permutations of  $\tau$ and $\mu,\sigma_{2}=(2,1,3)$ and $\nu_{2}=(2,1,4)$, respectively. Note that these permutations satisfy $\sigma^p = (\mu^p)^{-1}$, but have $\sigma^u \neq \mu^u$. If we write down the corresponding $R$ for both of these (using the definitions below equation \ref{V7}), we get: 
\begin{eqnarray}
\label{EQ11}
&& R_{1}	=	\mathrm{Perm}\left(\begin{array}{ccc}
M_{11}M_{12}^{*} & M_{12}M_{11}^{*} & |M_{13}|^{2}\\
M_{21}M_{22}^{*} & M_{22}M_{21}^{*} & |M_{23}|^{2}\\
M_{31}M_{32}^{*} & M_{32}M_{31}^{*} & |M_{33}|^{2}
\end{array}\right) \\
&& R_{2}	=	\mathrm{Perm}\left(\begin{array}{ccc}
M_{11}M_{12}^{*} & M_{12}M_{11}^{*} & |M_{14}|^{2}\\
M_{21}M_{22}^{*} & M_{22}M_{21}^{*} & |M_{24}|^{2}\\
M_{31}M_{32}^{*} & M_{32}M_{31}^{*} & |M_{34}|^{2}
\end{array}\right),
\end{eqnarray}
which only differ in their fixed points. Since these pick up the same phases, they have a nonzero covariance. 

In order to compute the variance of $c_{j}$, we now have two tasks. First, we must count the number of terms in equation \ref{V8} with satisfy the condition $\sigma^p = (\nu^p)^{-1}$. Secondly, we must compute the covariance between any two such terms. We will begin with the first task. To count the number of terms with which some given $R_{\tau,\sigma_{j}}$ has covariance, we must compute in how many ways we can assign $\sigma^u$ given $\sigma^p$. Observe that $\sigma$  is a permutation of $m$ elements, and that $j$ of those elements are set by $\sigma^p$. To assign the fixed points, of which there are $m-j$ by definition, we can chose from $n-j$ modes. The number of ways $ \#R$  in which we can assign the fixed points is therefore given by:
\begin{equation}
\label{EQ12a}
\#R=\left(\begin{array}{c}
n-j\\
m-j
\end{array}\right).
\end{equation}

It is interesting to note the extremal cases of equation (\ref{EQ12a}): if  $j=0$, we are in the term which corresponds to classical interference, and equation (\ref{EQ12a})  reduces to 
\[
\#R=\left(\begin{array}{c}
n\\
m
\end{array}\right),
\]
which is nothing more than the statement that all classiscal ways for photons to be transmitted through the system add without phase terms. In the case of  $j=m$,  $\#R=1$. The reason for this is that if there are no unpermuted parts in the permutation, there is also no way to assign them, and therefore the only permutation which has covariance with a given permutation is its inverse, just like in the lossless case. More generally, the number of terms which contribute covariance in equation \ref{V8} is a decreasing function of $j$. Therefore, as we will see later, terms with low $j$ dominate. Since these terms correspond to few-photon interference, this allows us to construct an efficient approximation to the probability distribution. 

To compute the variance of $c_{j}$ we have to estimate the covariance of $R$.  We
start with the observation that: 
\begin{equation}
\label{V9}
\mathrm{Cov}(R_{\tau,\sigma_{j}},R_{\mu,\nu_{j}})\leq\sqrt{\mathrm{Var}(R_{\tau,\sigma_{j}})\mathrm{Var( R_{\mu,\nu_{j}})}},
\end{equation}
where as was previously shown $\mathrm{Var}(R)<e\left(\begin{array}{c}
m\\
j
\end{array}\right)(m-j)!^{2}m!/2N^{2m}$ \cite{Renema2018}. Hence $\mathrm{Cov}(R_{\tau,\sigma_{j}},R_{\mu,\nu_{j}})\leq e\left(\begin{array}{c}
m\\
j
\end{array}\right)(m-j)!^{2 }m!/2N^{2m}$. This inequality is saturated when $\tau=\mu$ and $\sigma_{j}=\nu_{j}.$
In the case that $\sigma_{j}$ and $\nu_{j}$ are fully distinct in
their fixed points, we can compute (see also Ref. \cite{Renema2018}):
\begin{eqnarray}
\label{V10}
\mathrm{Cov}( R_{\tau,\sigma_{j}}, R_{\mu,\nu_{j}}) & = &  \langle  R_{\tau,\sigma_{j}} R_{\mu,\nu_{j}} \rangle -  \langle R_{\tau,\sigma_{j}}\rangle  \langle R_{\mu,\nu_{j}}\rangle \nonumber \\
 & = & \langle  R_{\tau,\sigma_{j}} R_{\mu,\nu_{j}}\rangle \nonumber \\
 & = &  \langle \sum_{\rho}\sum_{\rho'}S_{\tau,\sigma_{j},\rho}S_{\mu,\nu_{j}\rho'}\rangle \nonumber \\
 & = &   \langle \sum_{\rho}S_{\tau,\sigma_{j},\rho}S_{\mu,\nu_{j}\rho'}\rangle \nonumber  \\
 & = & \sum_{\rho} \langle S_{\tau,\sigma_{j},\rho}S_{\mu,\xi_{j}\rho'} \rangle \nonumber \\
 & = & \left(\begin{array}{c}
k\\
j
\end{array}\right)(k-j)!^{2}j!/2N^{2k},
\end{eqnarray}
where going from the third to fourth line we have used the fact that
there is phase cancellation if $\rho\neq\rho',$ and where in the
last line we have used the expression for $E(S)$ from the supplemental
material of \cite{Renema2018} (equation 18). Note that this lower bound
differs by a factor $1/e$ from the upper bound. In other words, if
two $R$ add in phase, their normalized covariance 
\[
1\leq \frac{\mathrm{Cov}(R_{\tau,\sigma_{j}},R_{\mu,\nu_{j}})}{\sqrt{\mathrm{Var(}R_{\tau,\sigma_{j}})}\sqrt{\mathrm{Var}(R_{\mu,\nu_{j}}})} \leq \frac{1}{e}.
\]

From Eqs. (\ref{V8}) and (\ref{V10}) we obtain:
\begin{eqnarray*}
\mathrm{Var}(c_{j}) & = & \sum_{\tau,\sigma_{j}}\mathrm{Var}( R_{\tau,\sigma_{j}})+\sum_{(\tau,\mu,\tau\neq\mu)}\sum_{(\sigma,\nu,\sigma\neq\nu)}\mathrm{Cov(} R_{\tau,\sigma_{j}}, R_{\mu,\nu_{j}}), \\
 & < & e\left(\begin{array}{c}
n-j\\
m-j
\end{array}\right)\left(\begin{array}{c}
n\\
m
\end{array}\right)D(m,m-j)(m-j)!^{2}\frac{m!}{N^{2m}},
\end{eqnarray*}
where $D(m,m-j)$ is the  rencontres number, which counts
the number of permutations  of $m$ elements with $m-j$ fixed points. Since    $D(m,m-j) = e^{-1}\left(\begin{array}{c}
m\\
m-j
\end{array}\right)j! + O(\frac{1}{(m-j)!})$, we obtain:  
\begin{equation}
\label{V11}
\mathrm{Var(c_{j})} < \left({n-j \atop m-j }\right)\left( {n\atop m  }\right)\left(\frac{m!}{N^m}\right)^2.
\end{equation}
Using this bound, the variance  in  Eq. (\ref{V4}) for  $P^\prime_m(q) \equiv \left({n\atop m} \right)^{-1}\sum_{j=0}^{k}x^j c_{j} $     is bounded as 
\begin{eqnarray}
\label{V12}
&& \mathrm{Var}(P_m(q) - P^\prime_m(q))  =  \left({n \atop m }\right)^{-2}\sum_{j = k+1}^m x^{2j}\mathrm{Var}(c_j) \nonumber\\
&& <\left({n \atop m}\right)^{-1} \left(\frac{m!}{N^m}\right)^2\sum_{j = k+1}^m x^{2j}  \left({n-j \atop m-j }\right) \nonumber\\
&& = \left(\frac{m!}{N^m}\right)^2\sum_{j = k+1}^m x^{2j}\frac{m(m-1) \ldots (m-j+1)}{n(n-1) \ldots (n-j+1)}  \nonumber\\
&& < \left(\frac{m!}{N^m}\right)^2 \sum_{j = k+1}^m \left(x^2\frac{m}{n}\right)^j < \left(\frac{m!}{N^m}\right)^2\frac{(x^2(m/n))^{k+1}}{1 - x^2(m/n)}.
\end{eqnarray}
where we have used the sum of the  geometric progression to obtain the last inequality. Note the important fact that on the r.h.s. of Eq. (\ref{V12}) we have the average probability squared $\langle P_m\rangle =m!/N^m$, thus our bound on the variance is in the form of Eq. (\ref{V4A}). Also note that the complex overlap between the internal states does  not change the result if we set  $x =   |\langle\phi_i|\phi_j\rangle|$.   Eq. (\ref{V12})  means that   the r.h.s. of Eq. (\ref{V4A}) is  bounded uniformly in $q$ by  $ \frac{\alpha_m^{k+1}}{(1 - \alpha_m)}\langle P_m\rangle^2$ with $\alpha_m \equiv x^2 m/n$. We can now state our first bound on the trace distance in Eq. (\ref{V2}). 

\section{Results}
\begin{theorem}
 The post-selected to $m$ detected photons  output probability distribution   of  a  uniformly  lossy boson sampling device    with $n$   partially distinguishable photons,  having a uniform overlap  $x$ of the internal states, $x \equiv |\langle\phi_i|\phi_j\rangle|$,  can be approximated by the   probability distribution  describing  a incoherent  mixture of $\left({m \atop k}\right)$ interferences of  $0\le k\le m$    partially distinguishable  (as above) photons    supplemented  with   $m-k$ classical particles    to the average (in a randomly chosen network)  trace distance error satisfying 
\begin{equation}
\langle d_m \rangle <    \sqrt{ \frac{\alpha_m^{k+1}}{1 - \alpha_m}} \equiv E(d_m),\quad  \alpha_m \equiv \frac{x^2m}{n}. 
\label{V13}
\end{equation}
\end{theorem}
Note that the bound in Eq. (\ref{V13}) depends on the strength of photon  losses, since the output probability distribution $P_m(q)$  in Eq. (\ref{V1}) involves $m$ input ports $\tau$  randomly chosen from the total number $n$.    

Theorem 1 states that one can approximate a lossy boson sampling with $n$ input photons and $m$ detected ones  by the  boson sampling with   $k$ coherent bosons and $m-k$ classical particles randomly distributed over $m$ ports chosen at random from $n$ input ones.      Obviously, it cannot have more quantum advantage  then the $k$-size boson sampling. 

Theorem 1 gives an upper bound   on  the trace distance between the output distribution and the approximation,   given  $k$ and $\alpha = x^2\frac{m}{n}$.  There is a similar result   for the bound  on the trace distance between the distributions with variable number of detected photons.   
\begin{theorem}
The full   output probability distribution   of  a  uniformly  lossy boson sampling device    with $n$   partially distinguishable photons,  having a uniform overlap  $x$ of the internal states, $x \equiv |\langle\phi_i|\phi_j\rangle|$,  can be approximated by the   probability distribution  describing  a incoherent mixture of $\left({n \atop k}\right)$ interferences of  $0\le k\le n$    partially distinguishable  (as above) photons  supplemented with   $n-k$ classical particles    to the average trace distance error satisfying  
\begin{equation}
\langle d\rangle  < 4e^{-2C^2}+  \sqrt{ \frac{\alpha^{k+1}}{1 - \alpha}}, \quad \alpha \equiv x^2\left(\eta + \frac{C}{\sqrt{n}}\right),
\label{V16}
\end{equation}
 \end{theorem}
for an arbitrary  $0<C < \sqrt{n}(1- \eta)$.  The proof of theorem 2 is as follows. We divide the summation in the trace distance between the whole distributions, i.e.,  
\begin{equation}
\label{S1}
d = \sum_{m=k+1}^{n}\left( {n \atop m}\right)\eta^m(1-\eta)^{n-m} d_m \equiv d^{(I_1)} +d^{(I_2)} +d^{(I_3)},
\end{equation}
into three parts according to the three domains $I_{1,2,3}$ of $m$: 
\begin{eqnarray}
\label{S2}
&& I_1 = \{ k+1\le m< n\eta - C\sqrt{n}\},\quad I_2 =  \{n\eta - C\sqrt{n}\le m \le n\eta + C\sqrt{n}\},\nonumber\\
&&  I_3= \{ n\eta - C\sqrt{n}< m\le n\},
\end{eqnarray}
where $C>0$.  Taking $C < \sqrt{n}(1- \eta)$, replacing the trace distance $d_m$ by its upper bound value $2$ (see Eq. (\ref{V2})), and using  Hoeffding's inequality for the binomial distribution, represented as a sum of $n$ i.i.d. Bernoulli trials $x_i\in \{0,1\}$,  \cite{Hoeffding} (Theorem 1 on p. 15), we have for the sum  in Eq. (\ref{S1}) over the domain $I_3$:
\begin{equation}
\label{S3}
d^{(I_{3})}  \le  2\mathrm{Prob}\left(\sum_{i=1}^nx_i - n\eta \ge C \sqrt{n}\right)\le 2\exp\left\{-2 n\left(\frac{C}{\sqrt{n}}\right)^2 \right\} = 2e^{-2C^2},
\end{equation}
and that  for the domain $I_1$, obtained by the symmetry of the binomial distribution: $i\to n-i$ and $\eta \to 1-\eta$ (i..e, by counting zeros instead of ones). The sum over the domain $I_2$ can be bounded using the fact that the upper bound on $d_m$, $E(d_m)$, is a growing function of $m$. We get 
\begin{equation}
\label{S4}
d^{(I_2)} \le \sum_{m\in I_2} \left( {n \atop m}\right)\eta^m(1-\eta)^{n-m}E(d_m) <  \sqrt{ \frac{\alpha^{k+1}}{1 - \alpha}}, \quad \alpha \equiv x^2\left(\eta + \frac{C}{\sqrt{n}}\right),
\end{equation}
where we have used that the sum of the  binomial distribution is bounded by $1$. Q.E.D.  

\textit{Theorem 2 reduces the problem of bounding the trace-distance of the  difference between the whole distributions to that for a fixed number of detected photons}.  Indeed,  by choosing $C(\varepsilon)= \sqrt{\ln(8/\varepsilon)/2}$  one converts the condition $\langle d\rangle \le \varepsilon$ on the trace distance between the whole distribution to the condition  $\langle d_m\rangle \le \varepsilon/2$   for a fixed number of detected photons $m\equiv n\eta + C(\varepsilon)\sqrt{n}$ (where the smallest bounding   integer  value  is implied).

Theorem 1 means the existence of a fixed-size  approximation to arbitrary imperfect boson sampling with non-ideal indistinguishability and/or losses.  Indeed, for  any given  trace distance error  $d_m\le \varepsilon\ll 1$ and any given   odds of failure $\delta\ll 1$  there is such $k(\alpha,\varepsilon\delta)$ that for any size $n\gg 1$ (the number of input photons to Boson Sampling) of   an imperfect setup for Boson Sampling, with $m$ out of $n$ photons detected  and the overlap $x$,  one can approximate its  output by an incoherent mixture of $\left({n \atop k}\right)$ interferences of  $0\le k\le n$    partially distinguishable  photons  supplemented by $n-k$ classical particles, where $k$ is at most 
\begin{equation}
\label{V17}
k(\alpha,\varepsilon\delta) =\left[\frac{\ln\left( \varepsilon \delta\frac{1-\alpha}{2}\right)}{\ln \alpha}\right] - 1, \quad \alpha = x^2\frac{m}{n}.
 \end{equation}
 To arrive at this conclusion we have used Markov's inequality for the trace distance $d$ of Eq. (\ref{V2})  and the bound on the average trace distance (\ref{V13}).